\documentclass{emulateapj}

\usepackage{graphicx,psfig,amssymb,verbatim,natbib}

\newcommand\lya{Ly$\alpha$}

\newcommand\wt{$w(\theta)$}

\newcommand\ip{$i_{775}$}

\newcommand\zp{$z_{850}$}
\newcommand\bp{$B_{435}$}

\newcommand\vp{$V_{606}$}

\newcommand\zs{$z\sim6$}

\shorttitle{Clustering of \ip\ dropout galaxies}
\shortauthors{Overzier, Bouwens, Illingworth \& Franx}

\begin{document}      
                      
\title{Clustering of \ip\ dropout galaxies at \zs\ in GOODS and the UDF\altaffilmark{1}}

\author{Roderik A. Overzier\altaffilmark{2}, Rychard J. Bouwens\altaffilmark{3}, Garth D. Illingworth\altaffilmark{3} and Marijn Franx\altaffilmark{2}}

\email{overzier@strw.leidenuniv.nl}

\altaffiltext{1}{Based on observations made with the NASA/ESA {\it Hubble Space Telescope}, which is operated by the 
Association of Universities for Research in Astronomy, Inc., under NASA contract NAS 5-26555.}

\altaffiltext{2}{Leiden Observatory, Postbus 9513, 2300 RA Leiden, Netherlands}

\altaffiltext{3}{UCO/Lick Observatory, University of California, Santa Cruz, CA 95064}

\begin{abstract}
We measured the angular clustering at $z\sim6$ from a large sample of \ip\ dropout galaxies (293 with \zp$\le$27.5 from GOODS 
and 95 with \zp$\le$29.0 from the UDF). Our largest and most complete subsample (having $L\gtrsim0.5L^*_{z=6}$) shows the presence of 
clustering at 94\% significance. For this sample we derive a (co-moving) correlation length of $r_0=4.5^{+2.1}_{-3.2}$ $h_{72}^{-1}$ Mpc 
and bias $b=4.1^{+1.5}_{-2.6}$, using an accurate model for the redshift distribution. 
No clustering could be detected in the much deeper but significantly smaller UDF, yielding $b<4.4$ (1$\sigma$).  
We compare our findings to Lyman break galaxies at $z\sim3-5$ at a fixed luminosity. 
Our best estimate of the bias parameter implies that \ip\ dropouts are hosted by dark matter halos having 
masses of $\sim10^{11}$ $M_\odot$, similar to that of \vp\ dropouts  
at $z\sim5$. We evaluate a recent claim  that at $z\gtrsim5$ star formation might have occurred more 
efficiently compared to that at $z=3-4$. This may provide an explanation for the very mild evolution observed in the 
UV luminosity density between $z=6$ and 3. Although our results are consistent with such a scenario, 
the errors are too large to find conclusive evidence for this. 
\end{abstract}



\section{Introduction}
\label{sec:intro}

The Advanced Camera for Surveys \citep[ACS;][]{ford98} aboard the {\it Hubble Space Telescope}   
has made the detection of star-forming galaxies at $z\sim6$ (\ip\ dropouts) relatively easy. 
The largest sample of \ip\ dropouts currently available \citep{bouwens05_z6} comes from the Great Observatories Origins 
Deep Survey \citep[GOODS;][]{giavalisco04_survey}, allowing the first quantitative analysis of galaxies 
only 0.9 Gyr after recombination \citep[][see also \citealt{shimasaku05,ouchi05}]{stanway03,bouwens03,yan04,dickinson04,malhotra05}. \citet{bouwens05_z6} found evidence for strong evolution of the luminosity function between 
$z\sim6$ and 3, while the (unextincted) luminosity density at $z\sim6$ is only $\sim0.8$ times lower than that at $z\sim3$. 
Some \ip\ dropouts have significant Balmer breaks, indicative of stellar populations older than 100 Myr 
and masses comparable to those of $L^*$ galaxies at $z\approx0$ \citep{eyles05,yan05}.

Through the study of the clustering we can address fundamental cosmological issues that cannot be answered from the study 
of galaxy light alone. The strength of clustering and its evolution with redshift 
allows us to relate galaxies with  
the underlying dark matter and study the bias. The two-point angular correlation function (ACF) has been used to measure the clustering of 
Lyman break galaxies (LBGs) at $z=3-5$ \citep[e.g.,][]{adelberger98,adelberger05,arnouts99,arnouts02,magliocchetti99,giavalisco01,ouchi01,ouchi04_r0,porciani02,hildebrandt05,allen05,kashikawa06}. 
LBGs are highly biased ($b\simeq2-8$), and this 
biasing depends strongly on rest frame UV luminosity and, to a lesser extent, on dust and redshift. 
The clustering statistics of LBGs have reached the 
level of sophistication that one can measure two physically different contributions. At small angular scales the 
ACF is dominated by the non-linear clustering of galaxies within single dark matter halos, whereas at large 
scales its amplitude tends to the ``classical'' clustering of galaxies residing in different 
halos \citep{ouchi05_smallscale,lee05}, as explained within the framework of the halo occupation distribution 
\citep[e.g.,][]{zehavi04,hamana04}. Understanding the clustering properties of galaxies at $z\sim6$ is 
important for the interpretation of 
``overdensities'' observed towards luminous quasars and in the field \citep{ouchi05,stiavelli05,wang05,zheng06} that could 
demarcate structures that preceded present-day massive galaxies and clusters \citep{springel05}.
Our aim here is to ``complete'' the census of clustering by extending it to the highest redshift regime 
with sizeable samples. In \S\S\ 2 and 3 we describe the sample,  
and present our measurements of the ACF. In \S\ 4 we discuss our findings. Throughout we use the cosmology ($\Omega_M$, $\Omega_\Lambda$, $h_{72}$, $n$, $\sigma_8$)$=$(0.27,0.73,1.0,1.0,0.9) with $H_0=72$ $h_{72}$ km s$^{-1}$ Mpc$^{-1}$. 

\begin{figure}[t]
\begin{center}
\includegraphics[width=\columnwidth]{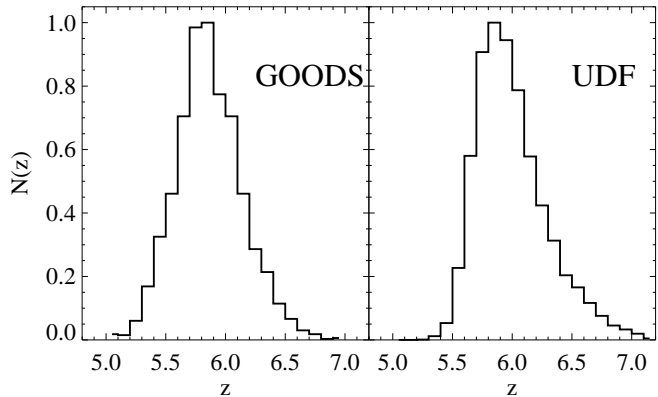}
\end{center}
\caption{\label{fig:nz}Redshift distributions of \ip\ dropouts in our GOODS ({\it left}) 
and UDF ({\it right}) selections \citep[estimated by projecting a complete UDF \bp\ dropout sample scaled to the sizes and colors as 
found for the \ip\ dropout sample to $z\sim5-7$; see][for details]{bouwens05_z6}. 
As a result of a more significant photometric scatter in \ip--\zp, the 
selection extends to lower redshifts in GOODS than it does for the UDF.}
\end{figure}

\section{Data}
\label{sec:data}

The present analysis is based on the sample of \ip\ dropouts described in detail by \citet{bouwens05_z6}. 
We used the ACS data from the GOODS v1.0 release, consisting of two spatially disjoint, $\sim160$ arcmin$^2$ fields. 
These data were processed with Apsis \citep{blakeslee03}, along with a 
substantial amount of overlapping data available from the Galaxy Evolution from Morphology and Spectral energy distributions \citep[GEMS;][]{rix04}, 
supernova searches (A. G. Riess et al. 2006 and S. Perlmutter et al. 2006, both in preparation), and Ultra Deep Field (UDF) NICMOS programs \citep[][]{thompson05}. 
The processed images were brought to a uniform signal-to-noise level by degrading the deeper parts of the area. The $10\sigma$ detection limit of the degraded data was 27.5 in \zp\ in a
$0\farcs2$ diameter aperture. 
We also used a deep sample of \ip\ dropouts selected from the UDF \citep{beckwith06}, covering one ACS pointing of $\sim11$ arcmin$^2$ with a 10$\sigma$ detection limit of 29.2. 

Objects were selected by requiring \ip--\zp$>$1.3, and \vp--\zp$>$2.8 or a non-detection (2$\sigma$) 
in \vp\ to exclude lower redshift interlopers. 
Point sources were removed based on high stellarity parameters $>$0.75. The estimated residual contamination due to 
photometric scatter, red interlopers, and stars is  $\sim$7\% to \zp$=$28.0, of which 2\% is due to stars \citep[see][for details]{bouwens05_z6}. 
The effective redshift distributions for GOODS and the UDF are shown in Figure \ref{fig:nz}. 
The effective rest-frame UV luminosity of the sample is 
$L\approx0.5L^*_{z=6}$ for \zp$\sim$27.5 \citep{bouwens05_z6}. 
Note that the luminosity is quite sensitive to redshift due to the Gunn-Peterson trough entering \zp\ at $z>6$, with $L^*_{z=6}$ corresponding to 
\zp$\sim$26.5 ($\sim$28) at $z=5.5$ ($z=6.5$).

\section{The angular correlation function}
\label{sec:results1}

We measured the ACF, \wt, defined as the excess probability of finding two sources in the solid angles $\delta\Omega_1$ and 
$\delta\Omega_2$ separated by the angle $\theta$, over that expected for a random Poissonian distribution \citep{peebles80}. 
We used the estimator $w(\theta)=[DD(\theta)-2DR(\theta)+RR(\theta)]/RR(\theta)$ of \citet{landy93}, where $DD(\theta)$, $DR(\theta)$ and 
$RR(\theta)$ are the number of pairs of sources with angular separations between $\theta$ and $\theta$$+$$\Delta\theta$ measured in 
the data, random, and data-random cross catalogs, respectively. We used 16 random catalogs containing $\sim100$ times more sources than 
in the data, but with
a similar angular geometry. Errors on \wt\ were bootstrapped \citep{ling86}. 
We assumed a power-law ACF of the form 
$w(\theta)$$=$$A_w\theta^{-\beta}$ and determined its amplitude, $A_w$, by fitting the function 
$w(\theta)$$=$$A_w\theta^{-\beta}$$-$$IC$. The integral constraint  
[$IC$$=$$\int\int w(\theta) d\Omega_1d\Omega_2/\Omega^2$, where $\Omega$ is the survey area] was $0.033A_w$ for GOODS and $0.074A_w$ for 
the UDF. We did not attempt to fit the slope of the ACF and assumed $\beta=0.6$ 
based on the results of \citet{lee05}. The ACF was fitted over the range  
10\arcsec--300\arcsec\ (10\arcsec--200\arcsec\ for the UDF), corresponding to roughly 0.4--10 $h_{72}^{-1}$ Mpc comoving at $z\sim6$. 
The lower value of 10\arcsec\ is larger than the virial radius of a $10^{12}$ $M_\odot$ halo to ensure that we are measuring 
the large-scale clustering (and not receiving a contribution at small scales from the subhalo component). 
Because the results of the fits are sensitive to the size of the bins used, we determined $A_w$  
using Monte Carlo simulations of the data.  
Finally, we note that if the contaminants to our samples ($\sim$7\% of the total) have a uniform distribution, the measured amplitude 
should be multiplied by $\sim$1.16 to yield the corrected clustering amplitude.

\begin{figure*}[t]
\begin{center}
\includegraphics[width=1.0\textwidth]{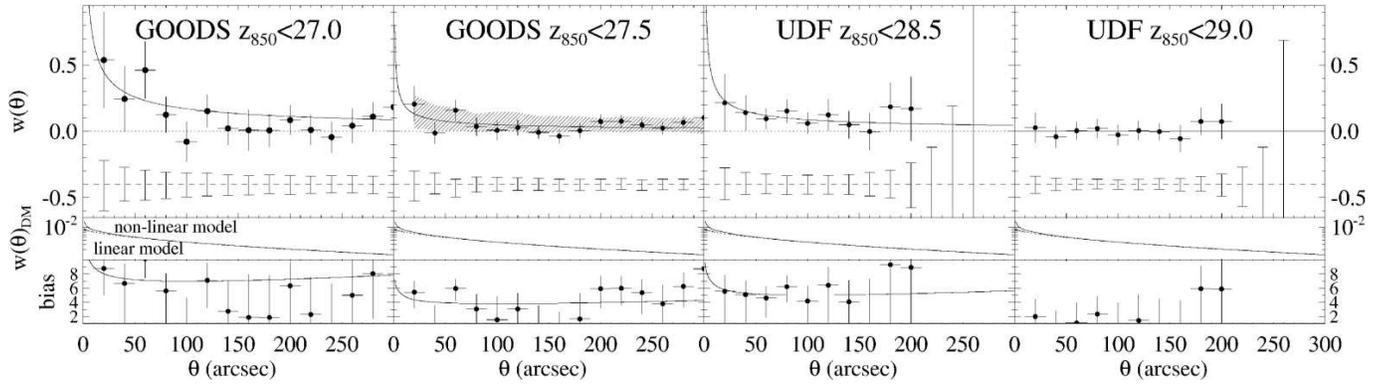}
\end{center}
\caption{\label{fig:measurements1}Top panels show the measurements ({\it points}) and the best-fit ACF ({\it solid line}) 
for the \ip\ dropout samples extracted from GOODS and the UDF. The slope was kept fixed at $\beta=0.6$. 
Shot noise expectations have been indicated by empty error bars (offset by --0.4 in the vertical direction for clarity). 
The middle and bottom panels show, respectively, the (non)linear clustering of dark matter \citep{pd96} 
and the bias at each $\theta$ ({\it points}) and for the best-fit ACF ({\it solid line}).
The hatched region indicates the $1\sigma$ range of \wt\ expected based on a simulated distribution with the same clustering amplitude as measured for the \zp$<$27.5 sample. }
\end{figure*}

\subsection{Results from GOODS and the UDF}

Figure \ref{fig:measurements1} ({\it top panels}) shows the GOODS and UDF ACFs for various limiting magnitudes. 
The results of the fits are given in Table \ref{tab:measurements}. 
For the three brightest subsamples (\zp$<$28.5) we measured a positive signal out to $\theta\sim1\arcmin$. 
In GOODS, we found $A_w=2.71\pm2.05$ for \zp$<$27.0, and $A_w\approx0.80\pm0.69$ for \zp$<$27.5. 
The analysis of the UDF is hampered by the relatively small number of sources available, owing to its $\sim30$ times smaller area, 
although its greater depth (1.5 mag) partially makes up for this lack of area. 
We found $A_w=1.40\pm1.64$ and $A_w=0.00\pm0.93$ for the \zp$<$28.5 and \zp$<$29.0 samples, respectively.

Because the objects were selected from data of uniform depth, signal in the ACF is unlikely to be caused by  
variations in the object surface density. Given the large errors on $A_w$, 
it is useful to ask whether the \wt\ observed at $\theta\lesssim1\arcmin$ could be the result 
of shot noise in a random object distribution. We created 1000 random distributions   
with the same geometry and the same number of points as our GOODS and UDF data, and calculated the ACF in each of the random samples.  
The mean and standard deviation at each $\theta$ is plotted in Figure \ref{fig:measurements1} ({\it top panels}, offset by --0.4 for clarity). 
We calculate the chance of reproducing the observed clustering in the random realizations, using the average \wt\ measured 
over the first four bins ($\theta<100\arcsec$) as a gauge of this clustering. This chance is 0.1\% for our \zp$<$27.0 sample, and 6\%, 10\% and 35\% for the fainter samples. 

\begin{deluxetable*}{lccccll}
\tablecolumns{7} 
\tablecaption{\label{tab:measurements}ACF and related physical quantities.} 
\tablewidth{0pt} 
\tablehead{\multicolumn{1}{c}{\zp} &  &\multicolumn{1}{c}{Area\tablenotemark{a}} & \multicolumn{1}{c}{$A_w$} & \multicolumn{1}{c}{$r_0$} & \\
 (mag)   & \multicolumn{1}{c}{$N$($<$$z$)}  & (arcmin$^2$)     & (arcsec$^\beta$)  &  ($h_{72}^{-1}$ Mpc) & \multicolumn{1}{c}{$b$ (30\arcsec)}}
\startdata 
\multicolumn{6}{c}{Enhanced\tablenotemark{b} GOODS Data}\\[0.5ex]
\hline
27.0 ...... & 172& $2\times160$    &  $2.71\pm2.05$ & $9.6^{+4.0}_{-5.6}$ & $7.5^{+2.5}_{-3.8}$ \\
27.5 ...... & 293& $2\times160$  &  $0.80\pm0.69$ & $4.5^{+2.1}_{-3.2}$ & $4.1^{+1.5}_{-2.6}$  \\[1ex]
\hline
\multicolumn{6}{c}{UDF Data}\\[0.5ex]
\hline
28.5 ...... & 52& $1\times11$    &  $1.40\pm1.64$ & $<10.2$ & $<8.0$    \\
29.0 ...... & 95& $1\times11$    &  $0.00\pm0.93$ & $<4.8$ & $<4.4$                 
\enddata 
\tablenotetext{a}{Approximate areal coverage that meets our $S/N$ requirements for \ip\ dropout selection \citep{bouwens05_z6}.}
\tablenotetext{b}{See \S\ 2 for details.}
\end{deluxetable*} 

Another test of the clustering signal detected in this sample was as follows. We used the formalism of \citet{soneira78} to  
create mock samples with a choice ACF in two dimensions. A $250\arcmin\times250\arcmin$ mock field with  
surface density similar to that of the \ip\ dropouts allowed us to mimic the measured $A_w$ to an accuracy of $\>98$\%, determined from a fit. 
Next, we randomly extracted 100 mock ``GOODS'' surveys and measured the mean $w(\theta)$ and its standard deviation 
using identical binning and fitting to that for the real samples. The result is indicated in Figure \ref{fig:measurements1} ({\it hatched region}) for our \zp$<$27.5 sample, which being our largest and most complete sample provides the most reliable constraint on this clustering. The simulation demonstrates that the amplitude of the observed \wt\ at $\theta\lesssim1\arcmin$ lies within $\lesssim1\sigma$ of the amplitudes predicted based on our model ACF, although 
the scatter in the expected amplitudes is large.

In the above analysis we restricted ourselves to clustering at $\theta$$\ge$10\arcsec. 
Our measurements also showed an excess of pair counts at $\theta$$<$10\arcsec. Upon closer inspection it was found that the excess was strictly
limited to $\theta$$<$5\arcsec, with $w(2\farcs5)$$\sim$$2.0\pm0.9$. The excess is consistent with an enhancement of \wt\ due to subhalo 
clustering at $\lesssim30$ kpc to $1.7\sigma$ confidence, but the exact amplitude cannot be determined accurately due to the small number of 
pairs (11 pairs at \zp$<$28.0). The excess is similar to that found for \lya\ emitters at $z=5.7$ \citep{shimasaku06}. While it is possible that the 
positive signal out to $\sim1$\arcmin\ 
is the result of strong subhalo clustering \citep[see][]{lee05,ouchi05}, the occurrence of such halos 
becomes increasingly rare with redshift, and by limiting the fits to $\theta\gtrsim$10\arcsec\ we minimized any 
contribution.  

\section{Derivation of cosmological quantities}
\label{sec:results2}

Although the uncertainties are large, we estimate the spatial correlation length ($r_0$) from $A_w$, using the Limber equation adopted 
for our cosmology and the redshift distributions of Figure \ref{fig:nz} (see Table \ref{tab:measurements}). 
The clustering was assumed to be fixed in comoving coordinates across the redshift range. 
We found $r_0$$=$4.5$^{+2.1}_{-3.2}$ $h_{72}^{-1}$ Mpc for the \zp$<$27.5 sample. 
At \zp$<$27, the best-fit value was found to be twice as high, $r_0$$=$9.6$^{+4.0}_{-5.6}$ $h_{72}^{-1}$ Mpc, but consistent 
with the fainter subsample within the errors. For the UDF samples, the best-fit values correspond to upper limits for the clustering amplitude 
of $\sim10$ and $\sim5$ $h_{72}^{-1}$ Mpc for \zp$<$28.5 and \zp$<$29, respectively. If we apply the contamination correction, $r_0$ increases by $\sim$10\%. 

We calculated the galaxy--dark matter bias, defined as $b(\theta)$$\equiv$$\sqrt{w(\theta)/w_{dm}(\theta)}$, 
where $w_{dm}(\theta)$ is the ACF of the dark matter as ``seen'' through our redshift window; $w_{dm}(\theta)$  
was calculated using the nonlinear fitting function of \citet{pd96} (Fig. \ref{fig:measurements1}, {\it middle panels}). 
In the bottom panels of Figure \ref{fig:measurements1} we have 
indicated the bias as a function of $\theta$ ({\it points}). Our best-fit ACF at \zp$<$27.5 
implies $b(\theta$$\sim$30$\arcsec)$$=$$4.1^{+1.5}_{-2.6}$ ({\it solid line}), 
bracketed by $b\sim8$ for the brightest GOODS sample and $b<4.4$ for the faintest UDF sample. The contamination correction yields values 
that are $\sim$5\% higher. 
\begin{figure}[t]
\begin{center}
\includegraphics[width=\columnwidth]{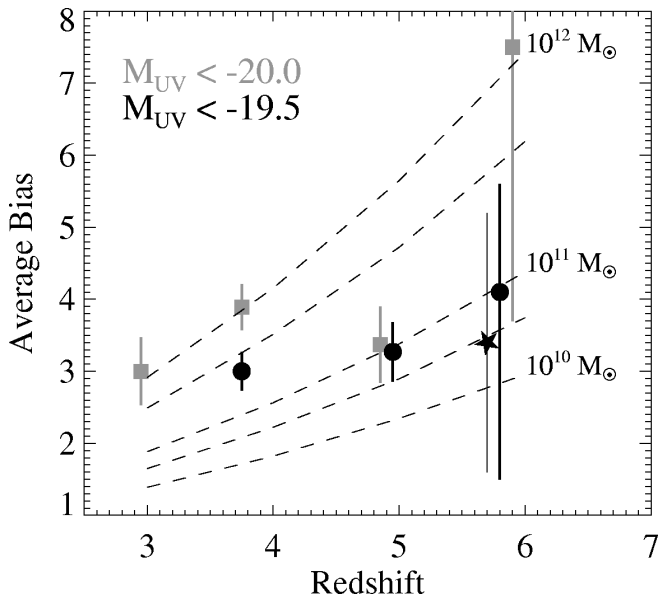}
\end{center}
\caption{\label{fig:measurements2}Bias of \ip\ dropouts compared to bias at $z=3-5$ from \citet{lee05}. Dashed lines indicate the 
bias of dark halos from \citet{sheth99} for $M_{halo}\ge10^{10}$, $5\times10^{10}$, $10^{11}$, $5\times10^{11}$, and $10^{12}$ $M_\odot$ ({\it bottom to top}). The relatively small halo mass inferred at $z=5$ compared to that at $z=4$ for objects with $M_z\lesssim-20$ ({\it squares}) cannot be confirmed at $z=6$ based on the present data. The best-fit halo mass inferred for objects at $z=6$ is consistent with the average halo mass of \vp\ dropouts at $z=5$ at a fixed luminosity of $M_z<-19.5$ ({\it circles}). The star indicates the bias of Ly$\alpha$ emitters at $z=5.7$ from \citet{ouchi05}.}
\end{figure}

It is important to evaluate how our results might be influenced by cosmic variance. Using \citet{somerville04}, 
we estimate $\sigma_v$$\sim$0.2 for GOODS and  
$\sigma_v$$\sim$0.5 for the UDF ($\sigma_v$ being the square root of the cosmic variance). 
Our best constraint on clustering at $z=6$ is therefore currently provided by the \zp$<$27.5 sample, 
given the relatively small variance, large sample size, and large completeness. Our best value for the bias of \ip\ dropouts is very similar to the 
bias of $b=3.4\pm1.8$ found for faint \lya\ emitters in the Subaru/{\it XMM-Newton} Deep Field \citep{ouchi05}.
An alternative method of estimating the bias is to directly compare the \ip\ dropout number density to the number density of dark halos 
at $z=6$. Assuming that the bias of the \ip\ dropouts corresponds to that of 
dark halos more massive than the average halo hosting them \citep{sheth99}, following \citet{somerville04} we predict that the average bias ranges from 
$b\approx5.5$ to $b\approx3.8$ for samples with limiting magnitudes from \zp$=$27 to \zp$=$29, with $\sim$5\% errors in these estimates  
due to the uncertainty in number density caused by cosmic variance. These values are largely in agreement with our measurements. The model predictions 
furthermore suggest that our best-fit value measured for the brightest GOODS sample (\zp$<$27.0) is likely spuriously high, 
given that the number density changes by not more than a factor of 2 over half a magnitude, giving only a modest increase in the bias compared to the 
\zp$<$27.5 sample. This GOODS sample thus adds very little additional constraints to the clustering at $z=6$. 
The UDF samples suffer from relatively small number statistics, as well as large cosmic variance. The clustering signal in the UDF is likely 
further diminished due to the strong luminosity dependence of clustering as seen at lower redshift \citep[e.g.,][]{kashikawa06}.

In Figure \ref{fig:measurements2} we compare our results to the work of \citet{lee05}, who found $b$$\approx$3.3$\pm$0.5 for faint 
\vp\ dropouts ($z\sim5$) also selected from GOODS. At \zp$\sim$27.5 we probe approximately the same 
rest-frame luminosity as their faintest (i.e., \zp$\le$27) \vp\ dropout sample ($M_z\lesssim-19.5$). To this limit, 
we measure a bias of $b=4.1^{+1.5}_{-2.6}$, suggesting an average halo mass of $\simeq10^{11}$ $M_\odot$,  
comparable to the average mass of halos hosting \vp\ dropouts. 

Interestingly, \citet{lee05} found that at slightly higher rest-frame luminosities ($M_z\lesssim-20$), the clustering of 
\vp\ dropouts is weaker than that of $U$ and \bp\ dropouts at $z=3-4$. It is hence inferred that the halo mass at $z=3-4$ is $\sim10$ times 
larger ($\sim$10$^{12}$ $M_\odot$) compared to that at $z=5$ (see Fig. 3). 
\citet{lee05} argued that star formation occurred more efficiently at higher redshifts ($z\sim5$) than it did 
at $z\sim3-4$, given that objects of comparable luminosity are found in less massive halos at $z\sim5$. 
Unfortunately, this result cannot be confirmed at $z\sim6$ using our brightest ($M_z\lesssim-20$) GOODS sample, 
given the large uncertainties in the bias and the associated halo mass. 
Also, the decrease in the effective halo mass from $z=4$ to 5 at $M_z\lesssim-19.5$   
is not as dramatic as observed at luminosities of $M_z\lesssim-20$, making it difficult to verify this result based on the present \ip\ dropout sample. 
However, we note that if a decrease in the star-forming efficiency with decreasing redshift is true 
(and can be confirmed for galaxies at $z\gtrsim6$), it would largely offset changes that are occurring in the mass function 
over this range. As such, this may provide at least a partial explanation for the mild evolution in the luminosity density from $z=6$ to 3.

In conclusion, we used the largest available sample of \ip\ dropouts to study clustering at $z\sim6$. We  
found a small signal, although its amplitude is not well constrained due to the large errors on the individual datapoints. 
The present analysis is therefore reminiscent of that performed at $z\sim3-5$ based on the original Hubble Deep Fields. 
The clustering of galaxies at $z\sim6$ will continue to be studied from deep, wide surveys \citep[e.g., see][]{ouchi05,shimasaku05,shimasaku06}. 
Although it might become possible in the near future to increase the size of our faint ACS samples by relaxing our current \ip\ dropout detection threshold, to perform an analysis at the same level of detail as currently performed at $z\sim5$ 
would require another six GOODS fields, for $\sim$1200 arcmin$^2$ in total. 

\acknowledgements

We thank the referee, Masami Ouchi, for a thorough report and helpful comments. ACS was developed under NASA contract NAS 5-32865, and this research
has been supported by NASA grant NAG5-7697.

\end{document}